\documentclass[preprintnumbers,notitlepage,superscriptaddress,nofootinbib]{revtex4-1}

\usepackage{amsfonts}
\usepackage{amsmath}
\usepackage{latexsym}
\usepackage{graphicx}
\usepackage{xcolor}
\usepackage{amssymb}
\usepackage{epsfig}
\usepackage{epstopdf}
\allowdisplaybreaks
\usepackage[caption=false]{subfig}
\usepackage[colorlinks=true, pdfstartview=FitV, linkcolor=red, citecolor=blue, urlcolor=blue]{hyperref}
\usepackage{MnSymbol}

\usepackage{qcircuit}
\usepackage{mathtools}
\DeclarePairedDelimiter\bra{\langle}{\rvert}
\DeclarePairedDelimiter\ket{\lvert}{\rangle}
\DeclarePairedDelimiterX\braket[2]{\langle}{\rangle}{#1 \delimsize\vert #2}

\newcommand{\be}{\begin{equation}}      
\newcommand{\ee}{\end{equation}}

\newcommand{\diff}{\mathrm{d}}

\newcommand{\p}{\partial}

\newcommand{\im}{\mathrm{i}}
\newcommand{\e}{\mathrm{e}}

\newcommand{\calT}{\mathcal{T}}

\newcommand{\calO}{\mathcal{O}}

\newcommand{\mt}{\mathrm{t}}

\newcommand{\bZ}{\mathbb{Z}}

\begin{document}

\preprint{RBRC-1321}

\title{Real-time chiral dynamics from a digital quantum simulation}

\author{Dmitri E. Kharzeev}
\email[]{dmitri.kharzeevATstonybrook.edu}

\affiliation{Department of Physics and Astronomy, Stony Brook University, Stony Brook, New York 11794-3800, USA}
\affiliation{Department of Physics, Brookhaven National Laboratory, Upton, New York 11973-5000}
\affiliation{RIKEN-BNL Research Center, Brookhaven National Laboratory, Upton, New York 11973-5000}

\author{Yuta Kikuchi}
\email[]{yuta.kikuchiATriken.jp}
\affiliation{RIKEN-BNL Research Center, Brookhaven National Laboratory, Upton, New York 11973-5000}

\begin{abstract}
The chiral magnetic effect in a strong magnetic field can be described using the chiral anomaly in the $(1+1)$-dimensional massive Schwinger model with a time-dependent $\theta$-term. We perform a digital quantum simulation of the model at finite $\theta$-angle and vanishing gauge coupling using an IBM-Q digital quantum simulator, and observe the corresponding vector current induced in a system of relativistic fermions by a global  {\it chiral quench} -- a sudden change in the chiral chemical potential or $\theta$-angle.
At finite fermion mass, there appears an additional contribution to this current that stems from the non-anomalous relaxation of chirality. Our results are relevant for the real-time dynamics of chiral magnetic effect in heavy ion collisions and in chiral materials, as well as for modeling high-energy processes at hadron colliders.
\end{abstract}
\maketitle

\renewcommand{\thefootnote}{\arabic{footnote}}
\setcounter{footnote}{0}

\setcounter{page}{1}


\section{Introduction}
\label{sec:intro}

Quantum theories possess a multi-dimensional Hilbert space that becomes very large for relativistic and/or many-body systems. This is why in addressing the real-time dynamics, the use of quantum simulations potentially provides an exponential advantage over the classical simulations in terms of required computational time and memory~\cite{Feynman:1981tf,Lloyd1073}.  Quantum simulations are also free from the sign problem that obstructs the use of Markov chain Monte-Carlo methods. 
Because of this, along with the rapid development of digital and analog quantum computers, quantum simulations  become a valuable source of information about the real-time behavior of relativistic and many-body systems~\cite{wallraff2004strong,majer2007coupling,Jordan:2011ne,Jordan:2011ci,Zohar:2012ay,Zohar:2012xf,Banerjee:2012xg,Banerjee:2012pg,Wiese:2013uua,Wiese:2014rla,Jordan:2014tma,Garcia-Alvarez:2014uda,Marcos:2014lda,Bazavov:2015kka,Zohar:2015hwa,Mezzacapo:2015bra,Dalmonte:2016alw,Zohar:2016iic,Martinez:2016yna,Bermudez:2017yrq,gambetta2017building,krinner2018spontaneous,Macridin:2018gdw,Zache:2018jbt,Zhang:2018ufj,Klco:2018kyo,Klco:2018zqz,Lu:2018pjk,Klco:2019xro,Lamm:2018siq,Gustafson:2019mpk,Klco:2019evd,Alexandru:2019ozf,Alexandru:2019nsa,Mueller:2019qqj,Lamm:2019uyc,Magnifico:2019kyj,Chakraborty:2019}.\\

The Chiral Magnetic Effect (CME) is a generation of electric current in an external magnetic field induced by the chiral asymmetry between the right- and left-handed fermions \cite{Fukushima:2008xe}, see \cite{kharzeev2014chiral,kharzeev2013strongly} for reviews and references. It is a non-equilibrium phenomenon stemming from the relaxation of chiral asymmetry via the chiral anomaly \cite{Bell:1969ts,Adler:1969gk}. In high-energy heavy ion collisions, the CME can reveal topological fluctuations in QCD matter \cite{kharzeev2006parity}. These fluctuations are akin to the electroweak sphalerons in the Early Universe that induce the baryon asymmetry. The experimental study of the effect is ongoing at Relativistic Heavy Ion Collider at BNL and the Large Hadron Collider at CERN, see \cite{kharzeev2016chiral} for review. This effect can also be studied in three-dimensional chiral materials (Dirac and Weyl semimetals) subjected to parallel electric and magnetic fields, and the CME has been observed in a Dirac semimetal ZrTe$_5$  \cite{li2016chiral} and other materials.\\

In a constant magnetic field and a given chiral chemical potential (the difference between the chemical potentials of the right- and left-handed fermions), the magnitude of the CME current is completely fixed by the chiral anomaly. However this is not so for a time-dependent magnetic field \cite{Kharzeev:2009pj}, or for a time-dependent chiral chemical potential. One of the most important effects that determine the real-time dynamics of the CME is the chirality flipping -- the transitions between the right- and left-handed fermions that are not related to the anomaly. The simplest mechanism producing such transitions arises from the finite masses of the fermions, since they induce non-conservation of the axial current. The mass effects are important for applications since the quarks in QCD are massive, and quasiparticles in many Dirac materials possess a finite gap.\\

In the limit of a strong magnetic field, the dynamics of chiral fermions becomes $(1+1)$-dimensional, since the fermions are frozen at the lowest Landau levels that are not degenerate in spin, and are thus chiral. Indeed, the real-time dynamics of CME, and of the {\it chiral magnetic wave} can be described \cite{Kharzeev:2010gd} within the $(1+1)$-dimensional QED, the Schwinger model \cite{Schwinger:1962tp}. The chiral anomaly relation in $(1+1)$-dimensions has the form 
\be\label{axial}
\partial_\mu J^\mu_5 = \frac{1}{\pi}\ E + 2\im m {\bar{\psi}}\gamma_5\psi,
\ee
where $J^\mu_5 = {\bar{\psi}}\gamma^{\mu}\gamma^5\psi$ is the axial current, $E$ is electric field, and $m$ is the fermion mass. 
The first term represents the chiral anomaly, and the second one is due to non-conservation of chirality induced by the masses of the fermions. It is well known \cite{Coleman:1975pw,Coleman:1976uz,Manton:1985jm}  that a background electric field $E_\text{cl}$ in the Schwinger model can be introduced through the $\theta$-angle: 
\be \label{electric}
E_\text{cl}=g\frac{\theta}{2\pi} .
\ee
The real-time dynamics of CME can thus be studied in the Schwinger model by considering the time-dependent 
$\theta$-angle. While the CME stems from the anomaly term, it is very important for applications to understand the role of the second term in (\ref{axial}) in the relaxation of the vector current. This is the goal of our work.\\

To accomplish this goal, we need to separate the effect of the anomaly from the effect caused by the masses of the fermions in (\ref{axial}). This can be achieved by first applying the chiral rotation $\psi\to\e^{\im\gamma_5\theta/2}\psi$ with $\theta = \theta(t)$ to the fermion fields $\psi$, and then by putting the coupling constant $g$ to zero.
In this limit of the theory, the time variation in $\theta$ does not induce a background electric field (see (\ref{electric})), but it does induce a chiral imbalance between the left- and right-handed fermions through the chiral chemical potential 
\be\label{chem}
\mu_5 = - \frac{\dot{\theta}}{2} .
\ee
Since the term describing the non-anomalous chirality flipping in (\ref{axial}) vanishes at $m=0$, in this limit of the theory the chiral chemical potential can induce the vector current only at finite fermion mass. 
In other words, a chirally imbalanced state with $\mu_5 \neq 0$ at $m=0$ cannot relax to a state with $\mu_5 = 0$.
This can be seen formally by observing that the Hamiltonian of the model in the chiral limit of $m=0$ commutes with the vector current operator, even at $\mu_5 \neq 0$. This means that a chiral imbalance indeed cannot induce a vector current in the chiral limit of massless fermions. \\

The situation changes when the fermions become massive. In this case, a chirally imbalanced state can relax to the true ground state with $\mu_5 = 0$, and generate a vector current during this relaxation process. It is clear that this is a real-time phenomenon, and we need to introduce a time-dependent chiral perturbation to study it. We will do this by subjecting the system to two different types of global (spatially independent) {\it chiral quenches}:  
\begin{enumerate}
\item Prepare the system in the state with $\theta = 0$ at times $t<0$. Then, starting at $t=0$, rotate the $\theta$-angle according to $\theta = - 2 \mu_5 t$, corresponding to a constant chiral chemical potential (\ref{chem}).
\item As before, prepare the system in the state with $\theta = 0$ at times $t<0$. Then abruptly change the $\theta$-angle 
at $t=0$ to a finite constant value corresponding to $\mu_5 = 0$; see \cite{Zache:2018cqq} for an earlier study of this quench type.
\end{enumerate}
We will refer to these two global quench protocols as the {\it $\mu_5$ quench} and the {\it $\theta$ quench}, respectively.
\section{Free fermion model at finite $\theta$-angle}

We choose the following basis for the gamma matrices,
\begin{align}
 \gamma_0 = Z
 \quad
 \gamma_1 = -\im Y,
 \quad
 \gamma_5 = \gamma_0\gamma_1 = -X,
\end{align}
where $X \equiv \sigma_x$, $Y \equiv \sigma_y$, $Z\equiv \sigma_z$ are the Pauli matrices.

The action of the massive Schwinger model with $\theta$ term in $(1+1)$-dimensional Minkowski space is
\begin{align}\label{action}
 S = \int\diff^2x\left[-\frac{1}{4} F^{\mu\nu} F_{\mu\nu} + \frac{g\theta}{4\pi}\epsilon^{\mu\nu}F_{\mu\nu} + \bar{\psi}(\im\gamma^\mu D_\mu-m)\psi\right],
\end{align}
with $D_\mu=\p_\mu-\im gA_\mu$.
Note that the gauge field $A_\mu$ and the coupling constant $g$ have mass dimensions 0 and 1, respectively. 
In what follows, we fix the gauge by putting $A_0=0$.
From the action (\ref{action}) the canonical momentum conjugate to $A_1$ can be read off as $\Pi = \dot{A}_1 - \frac{g\theta}{2\pi}$. The corresponding Hamiltonian is then given by
\begin{align}
 H &= \int \diff x \left[\frac{1}{2}\Big(\Pi+\frac{g\theta}{2\pi}\Big)^2
 +\bar{\psi}(\im \gamma_1D_1 +m)\psi\right],
\end{align}
with commutation relations $[A_1(x),\Pi(y)]=\im\delta(x-y)$, and $\{\psi(x),\bar{\psi}(y)\}=\gamma_0\delta(x-y)$.
Therefore, the term $g\theta/2\pi$ can be identified with a classical contribution to the total electric field $E = \dot{A}_1$ (in agreement  with (\ref{electric})), while $\Pi$ is the quantum contribution.

Upon the chiral transformation $\psi\to\e^{\im\gamma_5\theta/2}\psi$ and $\bar{\psi}\to\bar{\psi}\e^{\im\gamma_5\theta/2}$, the $\theta$-term is absorbed into the  phase of the fermion mass term,
\begin{align}
 S =  \int\diff^2x\Big[&-\frac{1}{4} F^{\mu\nu} F_{\mu\nu}
 + \bar{\psi}(\im\gamma^\mu D_\mu+\frac{\dot{\theta}}{2}\gamma_0\gamma_5-m\e^{\im\gamma_5\theta})\psi\Big],
\end{align}
where $\theta=\theta(t)$ is a time-dependent parameter.
We then decouple the gauge dynamics by taking the $g\to0$ limit to study the free fermionic theory with chiral chemical potential $\mu_5=- \dot{\theta}/2$ and chirally rotated mass term.
The resulting Hamiltonian is given by
\begin{align}\label{ham1}
 H &= \int \diff x \bar{\psi}\Big[\gamma_1\big(\im\p_1 -\frac{\dot{\theta}}{2}\big)+ m\e^{\im\gamma_5\theta}\Big]\psi .
\end{align}

Let us consider how the vector current is induced as a result of the chiral quench.
In our (1+1) dimensional system, the axial charge density $q_5(x)\equiv\bar{\psi}\gamma_5\gamma_0\psi(x)$ and the vector current density $j(x)\equiv\bar{\psi}\gamma_1\psi(x)$ are related by $q_5= - j$; the vector charge density $q(x)\equiv\bar{\psi}\gamma_0\psi(x)$ and axial current density $j_5(x)\equiv\bar{\psi}\gamma_5\gamma_1\psi(x)$ are identical, $q=j_5$. The Hamiltonian before the quench $H_0$ is obtained from (\ref{ham1}) by setting $\theta =0$ and ${\dot \theta}=-2\mu_5 = 0$. 
The full Hamiltonian after the quench is 
\begin{align}
\begin{split}
 H &= H_0 +\mu_5J + m\bar{\psi}(\e^{\im\gamma_5\theta}-1)\psi.
\end{split}
\end{align}
with $J=\int\diff xj(x)$.
The vector current is zero in the ground state before the quench $(t<0)$,
\begin{align}
\langle J \rangle_0 = 0,
\end{align}
where the expectation value $\langle\dots\rangle_0$ is taken with respect to the ground state of $H_0$.
In the massless case, the vector current is identically zero both before and after the quench,
\begin{align}
\langle \bar{\calT}[\e^{\im \int_0^t \diff tH}]J\, \calT[\e^{-\im \int_0^t \diff tH}]\rangle_0=0,
\end{align}
because $H_0$ and $H$ both commute with $J=-Q_5$.
$\calT$ ($\bar{\calT}$) indicates that the operator product is (anti-)time ordered.
With the fermion mass term included, $H$ and $J$ no longer commute,
\begin{align}
\label{comt}
 &[H,J] 
 = 2m\cos\theta\int\diff x\,\bar{\psi}\gamma_5\psi(x)+2\im m\sin\theta\int\diff x\,\bar{\psi}\psi(x).
\end{align}
Hence, the current can take a finite value. Indeed, the current does not vanish and at short times behaves as (Appendix~\ref{app:commutation})
\begin{align}
\label{short}
\begin{split}
 \langle \bar{\calT}[\e^{\im \int_0^{t'}\diff tH}]J\, \calT[\e^{-\im \int_0^{t'}\diff tH}]\rangle_0
 &=\im\int_0^t\diff t_1\langle [H(t_1),J]\rangle_0 -\int_0^t\diff t_1\int_0^{t_1}\diff t_2\langle [H(t_2),[H(t_1),J]]\rangle_0
 + \calO(t^3)
 \\
 &=\Big(-2m\int_0^t\diff t_1\sin\theta(t_1)-2m\int_0^t\diff t_1\int_0^{t_1}\diff t_2\dot{\theta}(t_2)\cos\theta(t_1)\Big)\int\diff x\langle\bar{\psi}\psi\rangle_0
 \\
 &-4m\int_0^t\diff t_1\int_0^{t_1}\diff t_2\sin\theta(t_1)\int\diff x\langle\bar{\psi}\gamma_5\p_1\psi\rangle_0
 + \calO(t^3).
\end{split}
\end{align}

\subsection*{Spin Hamiltonian of the lattice fermion model}

Let us now set up this problem in the lattice form suitable for a digital quantum simulation.
We impose a periodic boundary condition, where the $0$th and $N$th sites are identified and $N$ is an even integer. 
The staggered Hamiltonian is~\cite{casher1974vacuum,Susskind:1976jm}
\begin{align}
\begin{split}
 H &= -\im w\sum_{n=0}^{N-1}\left[\chi_{n}^\dag\chi_{n+1}-\chi_{n+1}^\dag\chi_{n}\right] 
 -\frac{\dot{\theta}}{4}\sum_{n=1}^{N-1}\left[\chi_{n}^\dag\chi_{n+1}+\chi_{n+1}^\dag\chi_{n}\right] 
 \\
 &+ m\cos\theta\sum_{n=0}^{N-1}(-1)^n\chi_n^\dag\chi_n
 + \im \frac{m}{2}\sin\theta\sum_{n=0}^{N-1}(-1)^n\left[\chi_{n}^\dag\chi_{n+1}-\chi_{n+1}^\dag\chi_{n}\right],
\end{split}
\end{align}
where $a$ is the lattice spacing and $w=(2a)^{-1}$ (see Appendix~\ref{app:staggered} for details).
For the purpose of quantum simulation, we apply the Jordan-Wigner transformation~\cite{Jordan:1928wi},
\begin{align}
\label{eq:JW}
 \chi_n =\frac{X_n-\im Y_n}{2} \prod_{i=0}^{n-1}(-\im Z_i), \qquad
 \chi_n^\dag = \frac{X_n+\im Y_n}{2}\prod_{i=0}^{n-1}\im Z_i,
\end{align}
which leads to the desired spin Hamiltonian:
\begin{align}
\label{eq:spin_ham}
\begin{split}
 &H = H_1 + H_2 + H_3 + H_4 + H_5,
 \\
 &H_1 =\frac{1}{2}\sum_{n=0}^{\frac{N}{2}-1}\left(w-\frac{m}{2}(-1)^n\sin\theta\right)
 \left[X_{2n}X_{2n+1}+Y_{2n}Y_{2n+1}\right],
 \\
 &H_2 =\frac{1}{2}\sum_{n=1}^{\frac{N}{2}-1}\left(w-\frac{m}{2}(-1)^n\sin\theta\right)
 \left[X_{2n-1}X_{2n}+Y_{2n-1}Y_{2n}\right]
 \\
 &+\frac{(-1)^{\frac{N}{2}}}{2}\left(w-\frac{m}{2}(-1)^{N-1}\sin\theta\right)
 \left[X_{N-1}X_{0}+Y_{N-1}Y_{0}\right]\prod_{i=1}^{N-2}Z_i,
 \\
 &H_3= -\frac{\dot{\theta}}{8}\sum_{n=0}^{\frac{N}{2}-1}
 \left[X_{2n}Y_{2n+1}-Y_{2n}X_{2n+1}\right],
 \\
 &H_4= -\frac{\dot{\theta}}{8}\sum_{n=1}^{\frac{N}{2}-1}
 \left[X_{2n-1}Y_{2n}-Y_{2n-1}X_{2n}\right]
 -\frac{(-1)^{\frac{N}{2}}\dot{\theta}}{8} \left[X_{N-1}Y_{0} - Y_{N-1}X_{0}\right]\prod_{i=1}^{N-2}Z_i,
 \\
 &H_5= \frac{m\cos\theta}{2}\sum_{n=0}^{N-1}(-1)^nZ_n.
\end{split}
\end{align}
It should be stressed that each $H_i$ does not violate the particle number conservation. This decomposition enables us to carry out the Suzuki-Trotter decomposition~\eqref{eq:suzuki_trotter} with the particle number conservation preserved exactly.

We are interested in computing the spacial average of vector current, which, in terms of spin operators, is expressed as
\begin{align}
\label{eq:J_average}
\begin{split}
 &\bar{J} = \bar{J}_1 + \bar{J}_2,
 \\
 &\bar{J}_1 = \frac{w}{2N} \sum_{n=0}^{\frac{N}{2}-1}(X_{2n}Y_{2n+1} - Y_{2n}X_{2n+1}),
 \\
 &\bar{J}_2 = \frac{w}{2N} \sum_{n=1}^{\frac{N}{2}-1}(X_{2n-1}Y_{2n} - Y_{2n-1}X_{2n})
 +\frac{(-1)^{\frac{N}{2}}w}{2N} (X_{N-1}Y_{0} - Y_{N-1}X_{0})\prod_{i=1}^{N-2}Z_i.
\end{split}
\end{align}

We note that the relation \eqref{comt} and \eqref{short} hold for corresponding operators defined on the lattice without taking a continuum limit.

\section{Real-time simulation}

We implement the quantum simulation for the global chiral quench using a IBM~Q digital quantum simulator in three following steps: initial state preparation, real-time evolution, and measurement.

\subsection{Initial state preparation}

We prepare the ground state, $|\Psi(0)\rangle$, of the Hamiltonian~\eqref{eq:spin_ham} at $\theta=0$ and ${\dot \theta}=0$ using a python package for exact diagonalization, QuSpin~\cite{weinberg2017quspin}.\footnote{Adiabatic state preparation of the massive Schwinger model at finite $\theta$-angle was studied in detail in~\cite{Chakraborty:2019} by carefully taking thermodynamic and continuum limits of the lattice Schwinger model.}
This state is used as the initial state in our studies of real-time chiral dynamics.

\subsection{Time evolution with the Hamiltonian at $\theta\neq0$}\label{subsec:evo}

The time-evolution operator is applied to the initial state,
\begin{align}
|\Psi(t)\rangle = \calT[\e^{-\im \int_0^t \diff t'H(t')}]|\Psi(0)\rangle.
\end{align}
We employ the trapezoidal discretization of integration and the Suzuki-Trotter decomposition in order to approximate the time-evolution operator, $\calT[\e^{-\im \int_0^t \diff t'H(t')}]$, by elementary unitary gates.
Applying the trapezoidal rule we approximate the integral by the sum,
$\sum_{s=0}^{S}\tilde{H}^s:= (H(0)+H(t))/2 + \sum_{s=1}^{S-1}H(st/S)
 = \int_0^t \diff t'H(t') + \calO(t^2/S^2)$.
Then, the Suzuki-Trotter decomposition is done as follows:
\begin{align}
\label{eq:suzuki_trotter}
\begin{split}
 &\prod_{s=0}^{S}\e^{-\im(\tilde{H}^s_5+\tilde{H}^s_4+\tilde{H}^s_3+\tilde{H}^s_2+\tilde{H}^s_1)\frac{t}{S}}
 \\
 &=\e^{\im \tilde{H}^{S+1}_{2} \frac{t}{2S}}
 \prod_{s=0}^{S}
 \big(\e^{-\im (\tilde{H}^{s+1}_{2}+\tilde{H}^s_{2}) \frac{t}{2S}}
 \e^{-\im \tilde{H}^s_{1} \frac{t}{2S}}
 \e^{-\im \tilde{H}^s_{3} \frac{t}{2S}}
 \e^{-\im \tilde{H}^s_5 \frac{t}{2S}}
 \e^{-\im \tilde{H}^s_{4}\frac{t}{S}}
 \e^{-\im \tilde{H}^s_5 \frac{t}{2S}}
 \e^{-\im \tilde{H}^s_{3} \frac{t}{2S}}
 \e^{-\im \tilde{H}^s_{1} \frac{t}{2S}}\big)
 \e^{-\im \tilde{H}^0_{2} \frac{t}{2S}} 
 +\calO(t^3/S^2),
\end{split}
\end{align}
where $\tilde{H}_i^s:=\tilde{H}_i(st/S)$.
The operator product is understood to be time ordered.
Hence, the total decomposition error is $\calO(t^3/S^2)$, which we call the Trotter error.
The circuit implementation of each unitary is given in Appendix~\ref{app:circuit}.

Here, we consider two protocols for the quench described in Sec.~\ref{sec:intro}:
\begin{enumerate}
\item  $\mu_5$ quench:\\
At $t=0$ we abruptly turn on the chiral chemical potential $\mu_5$ and keep it constant during the subsequent evolution.
\item $\theta$ quench:\\
At $t=0$ we abruptly change the value of the $\theta$-angle, that introduces the pulse in the chiral chemical potential $-\frac{\theta}{2}\delta(t)$. 
\end{enumerate}

\begin{figure}[t]
\centering
\includegraphics[scale=0.5]{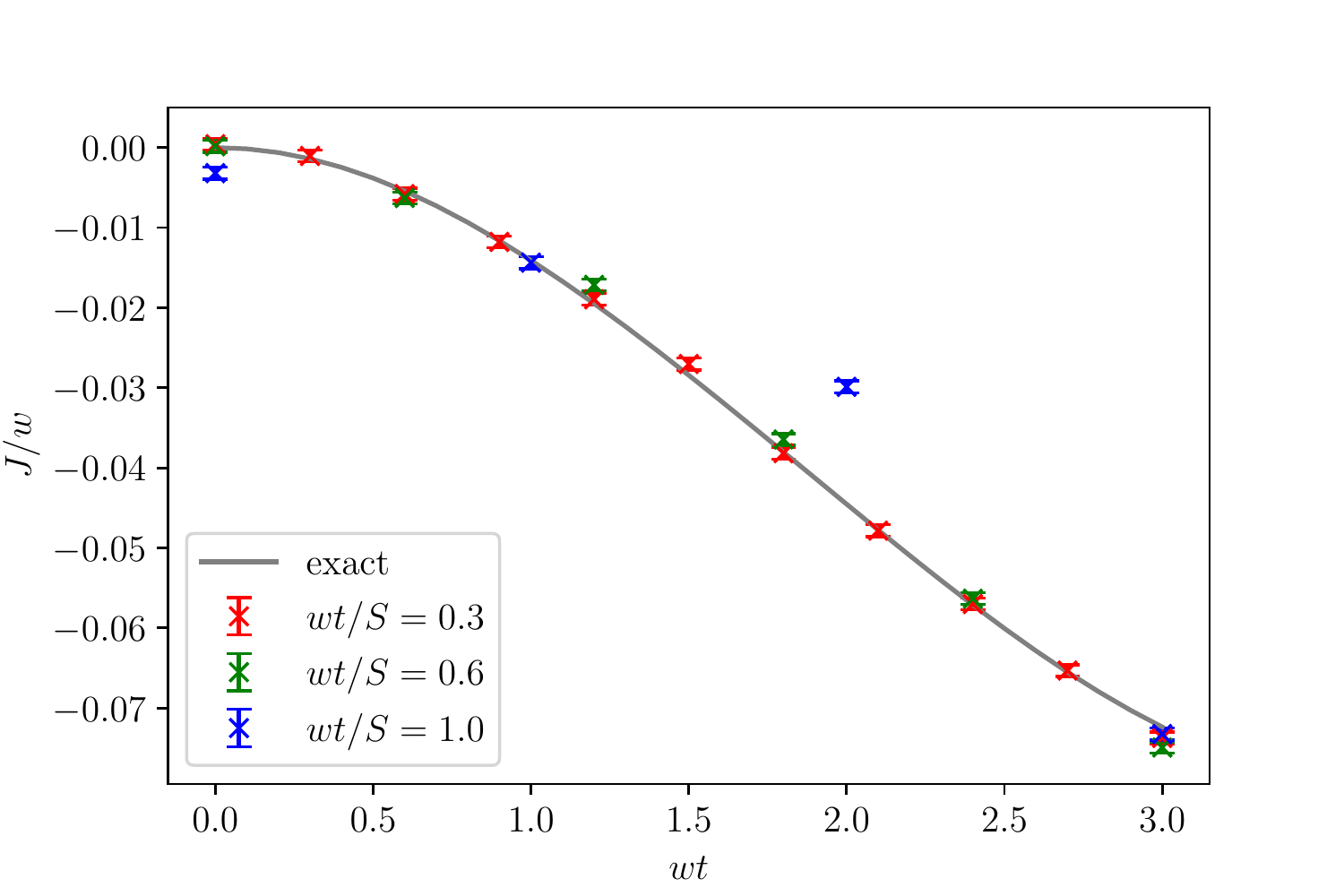}
\caption{The $t/S$-dependence of the vector current following the $\mu_5$ quench, with lattice size $N=12$, fermion mass $M=0.1w$, and chiral chemical potential $\mu_5=0.2w$. The error bar for each data point represents the statistical error which is estimated by performing 100000 measurements. The solid lines are exact results.
The discretization error becomes significant for $wt/S=1.0$.}
\label{fig:dTdep}
\end{figure}

The unitary evolution with the use of the Suzuki-Trotter decomposition may be interpreted as the Hamiltonian evolution with the shadow Hamiltonian $H_\text{sh}$, defined by
\begin{align}
\label{eq:Hshadow}
 \e^{-\im H_\text{sh}t} :=\e^{\im \tilde{H}^{S+1}_{2} \frac{t}{2S}}
 \prod_{s=0}^{S}
 \big(\e^{-\im (\tilde{H}^{s+1}_{2}+\tilde{H}^s_{2}) \frac{t}{2S}}
 \e^{-\im \tilde{H}^s_{1} \frac{t}{2S}}
 \e^{-\im \tilde{H}^s_{3} \frac{t}{2S}}
 \e^{-\im \tilde{H}^s_5 \frac{t}{2S}}
 \e^{-\im \tilde{H}^s_{4}\frac{t}{S}}
 \e^{-\im \tilde{H}^s_5 \frac{t}{2S}}
 \e^{-\im \tilde{H}^s_{3} \frac{t}{2S}}
 \e^{-\im \tilde{H}^s_{1} \frac{t}{2S}}\big)
 \e^{-\im \tilde{H}^0_{2} \frac{t}{2S}}.
\end{align}
The shadow Hamiltonian $H_\text{sh}$ differs from the original Hamiltonian \eqref{eq:spin_ham} by the Trotter error.
Since we use the exact ground state of \eqref{eq:spin_ham} as an initial state, the Hamiltonian evolution $H_\text{sh}$ introduces an additional quench due to the Trotter error in addition to the $\mu_5$ or $\theta$ quench. This artificial quench accounts for a small oscillation of $\calO(t^3/S^2)$ in the physical quantities such as vector current as shown in Fig.~\ref{fig:dTdep}.

\subsection{Evaluation of the vector current}\label{subsec:ev}

Having obtained the $|\Psi(t)\rangle$ at time $t$, we evaluate the expectation value of the vector current density $\langle\bar{J}\rangle:=\bra{\Psi(t)} \bar{J} \ket{\Psi(t)}$, where $\bar{J}$ is given by \eqref{eq:J_average}.
Since the operators $\bar{J}_1$ and $\bar{J}_2$ do not commute with each other, we carry out two independent measurements to read off the expectation value of $\bar{J}$.

Let us first find a way to calculate $\langle\bar{J}_1\rangle$.
We note the following identities:
\begin{align}
\begin{split}
 &(HS^\dag)_iCX_{ij}(X_iY_j)CX_{ij}(SH)_i
 =Z_iZ_j,
 \\
 &(HS^\dag)_iCX_{ij}(Y_iX_j)CX_{ij}(SH)_i
 =Z_iI_j,
\end{split}
\end{align}
where $H_i$ and $S_i$ are Hadamard and $\pi/4$ gates acting on the $i$th qubit. $CX_{ij}$ is a CNOT gate with the $i$th and $j$th qubits being control and target qubits, respectively.
These identities imply that the $Z_i$- and $Z_j$-measurements on a state $(SH)_iCX_{ij}\ket{\psi}$ yield the data needed to compute $\langle X_iY_j\rangle$ and $\langle Y_iX_j\rangle$.
The measurement is performed by the following circuit:
\begin{align}
\begin{array}{c}
\Qcircuit @C=2mm @R=2mm{
\lstick{j}&\targ & \qw &\qw &\meter
\\
\lstick{i}&\ctrl{-1} & \gate{S^\dag} & \gate{H}&\meter
}
\end{array}
\end{align}
Given the counts of the digits $(d_i,d_j)$ from the above measurements, we can extract the operator expectation values,
\begin{align}
\label{eq:XYexp}
\begin{split}
 \langle X_iY_j\rangle &= \sum_{(d_i,d_j)}(1-2d_i)(1-2d_j)\frac{\text{counts}_{d_i,d_j}}{\text{shots}},
 \\
 \langle Y_iX_j\rangle &= \sum_{(d_i,d_j)}(1-2d_i)\frac{\text{counts}_{d_i,d_j}}{\text{shots}},
\end{split}
\end{align}
which allows us to compute $\langle \bar{J}_1\rangle$.

The same argument applies to $\bar{J}_2$ except for the operators $X_{N-1}Y_{0}\prod_{i=1}^{N-2}Z_i$ and $Y_{N-1}X_{0}\prod_{i=1}^{N-2}Z_i$. 
We use the identities
\begin{align}
\begin{split}
 &\Big(\prod_{n=1}^{\frac{N}{2}}(HS^\dag)_{2n-1}CX_{2n-1,2n}\Big)
 \Big(X_{N-1}Y_{0}\prod_{i=1}^{N-2}Z_i\Big)
 \Big(\prod_{m=1}^{\frac{N}{2}}CX_{2m-1,2m}(SH)_{2n-1}\Big)
 =Z_0\Big(\prod_{i=1}^{\frac{N}{2}-1}Z_{2i}\Big)Z_{N-1},
 \\
 &\Big(\prod_{n=1}^{\frac{N}{2}}(HS^\dag)_{2n-1}CX_{2n-1,2n}\Big)
 \Big(Y_{N-1}X_{0}\prod_{i=1}^{N-2}Z_i\Big)
 \Big(\prod_{m=1}^{\frac{N}{2}}CX_{2m-1,2m}(SH)_{2n-1}\Big)
 =I_0\Big(\prod_{i=1}^{\frac{N}{2}-1}Z_{2i}\Big)Z_{N-1},
\end{split}
\end{align}
where the $N$th site is identified with the $0$th site.
Therefore, provided the counts of the digit $(d_0,\dots,d_{N-1})$, we can extract the operator expectation values,
\begin{align}
\begin{split}
 \langle X_{N-1}Y_{0}\prod_{i=1}^{N-2}Z_i\rangle 
 &= \sum_{(d_0,\dots,d_{N-1})}\Big(\prod_{i=0}^{\frac{N}{2}-1}(1-2d_{2i})\Big)(1-2d_{N-1})\frac{\text{counts}_{d_0,\dots,d_{N-1}}}{\text{shots}},
 \\
 \langle Y_{N-1}X_{0}\prod_{i=1}^{N-2}Z_i\rangle 
 &= \sum_{(d_0,\dots,d_{N-1})}\Big(\prod_{i=1}^{\frac{N}{2}-1}(1-2d_{2i})\Big)(1-2d_{N-1})\frac{\text{counts}_{d_0,\dots,d_{N-1}}}{\text{shots}}.
\end{split}
\end{align}
These, combined with \eqref{eq:XYexp}, provide all the ingredients required to compute $\langle \bar{J}_2\rangle$.

\section{Results and discussion}

\begin{figure}[t]
\centering
\begin{minipage}{.49\textwidth}
\subfloat[]{
\includegraphics[scale=0.5]{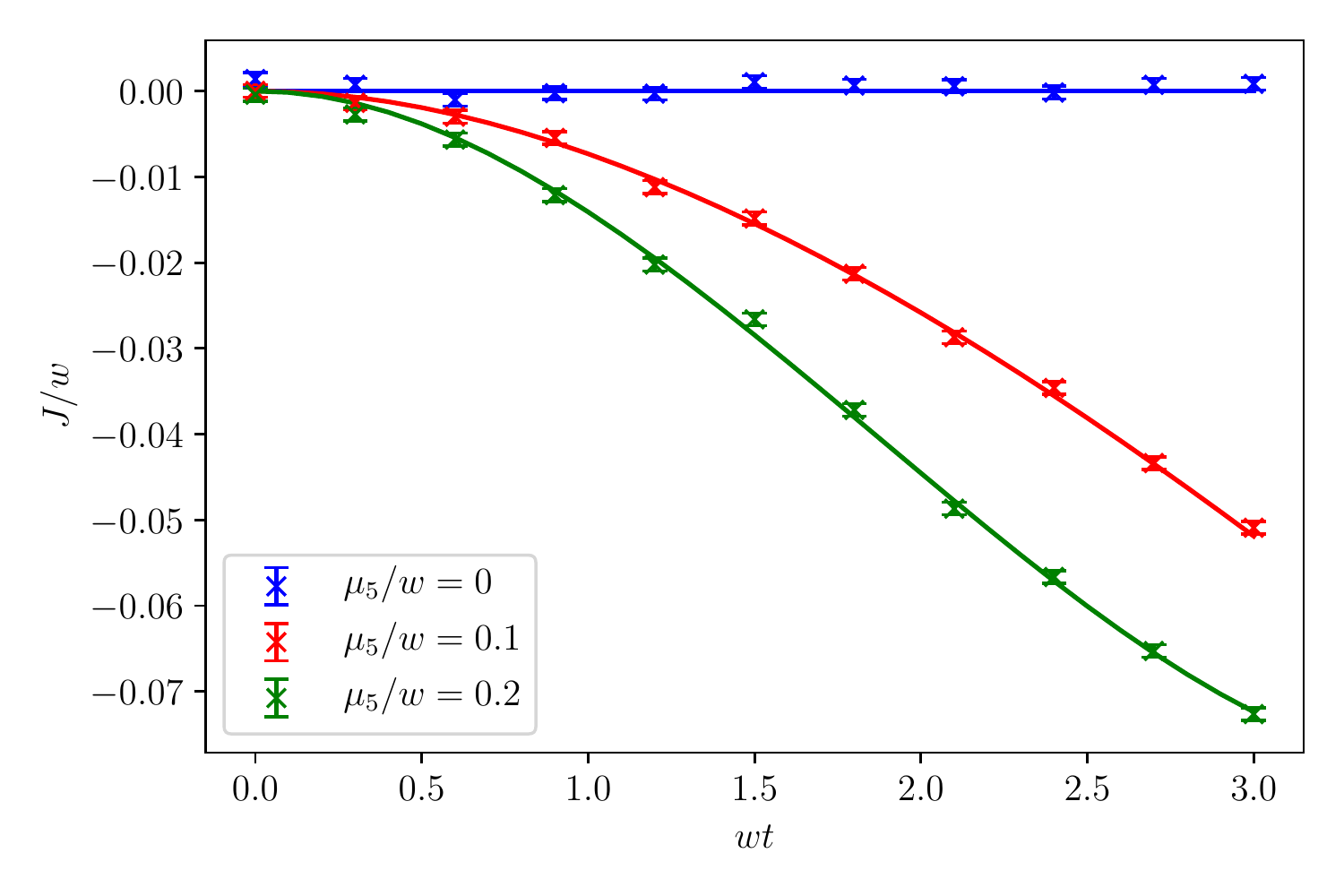}
\label{fig:current1a}
}\end{minipage}\
\begin{minipage}{.49\textwidth}
\subfloat[]{
\includegraphics[scale=0.48]{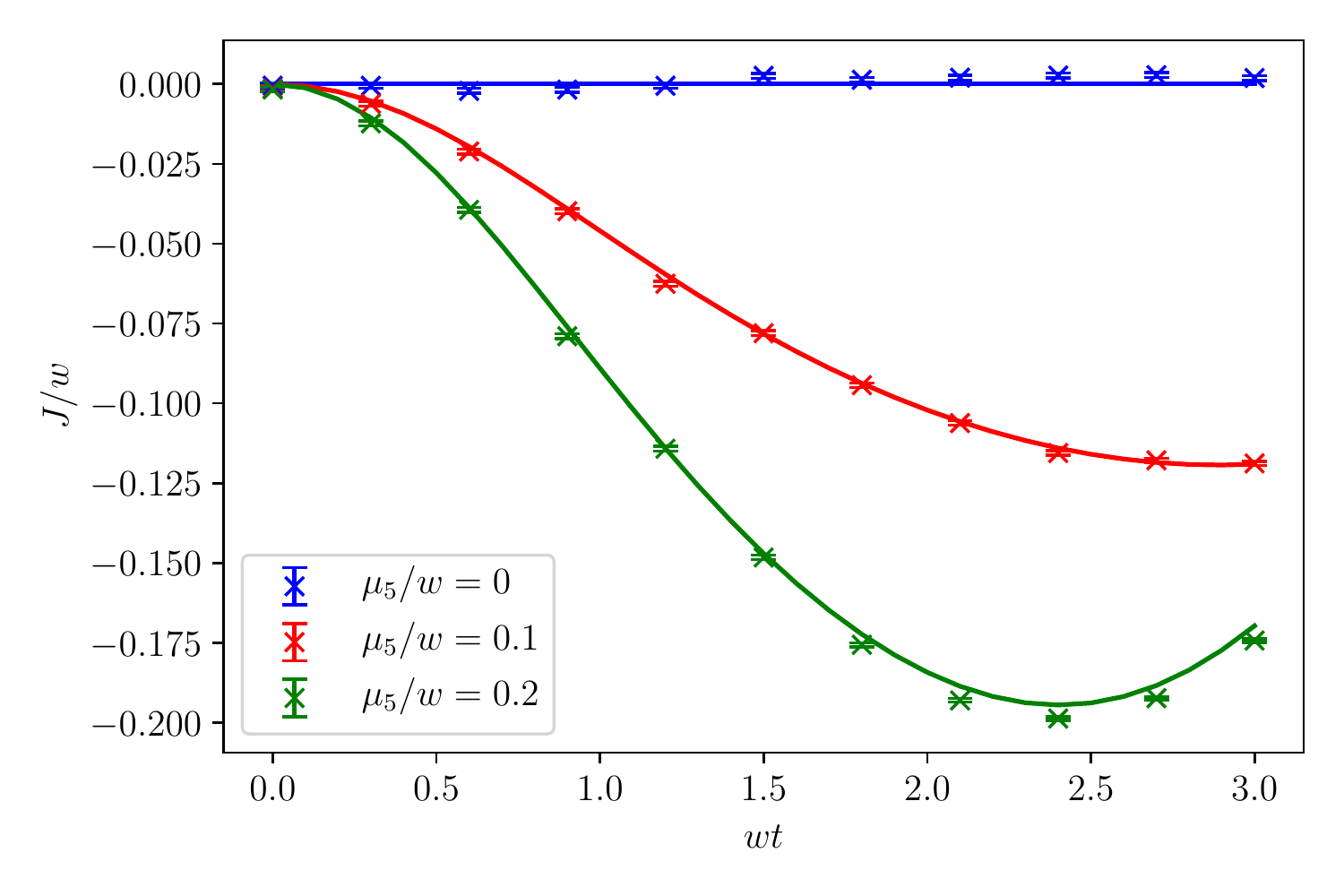}
\label{fig:current1b}
}\end{minipage}
\caption{The vector current densities after the $\mu_5$ quench with lattice size $N=12$ and temporal lattice spacing $wt/S=0.3$. The fermion mass is set to be (a) $M=0.1w$ and (b) $M=0.5w$. The error bar for each data point represents the statistical error which are estimated by performing 100000 measurements. The solid lines are exact results.}
\label{fig:current1}
\end{figure}

\begin{figure}[t]
\centering
\begin{minipage}{.49\textwidth}
\subfloat[]{
\includegraphics[scale=0.5]{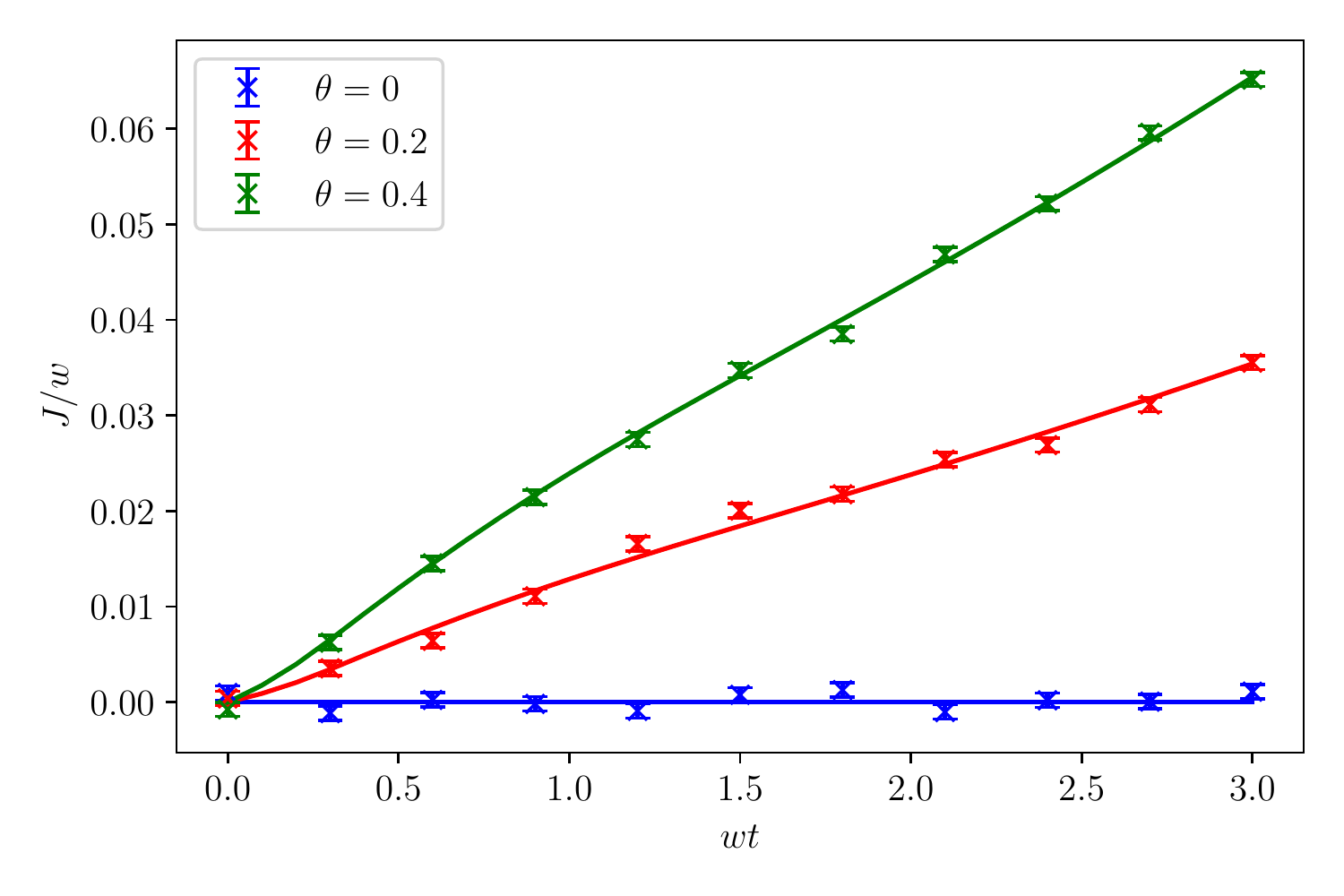}
\label{fig:current2a}
}\end{minipage}\
\begin{minipage}{.49\textwidth}
\subfloat[]{
\includegraphics[scale=0.48]{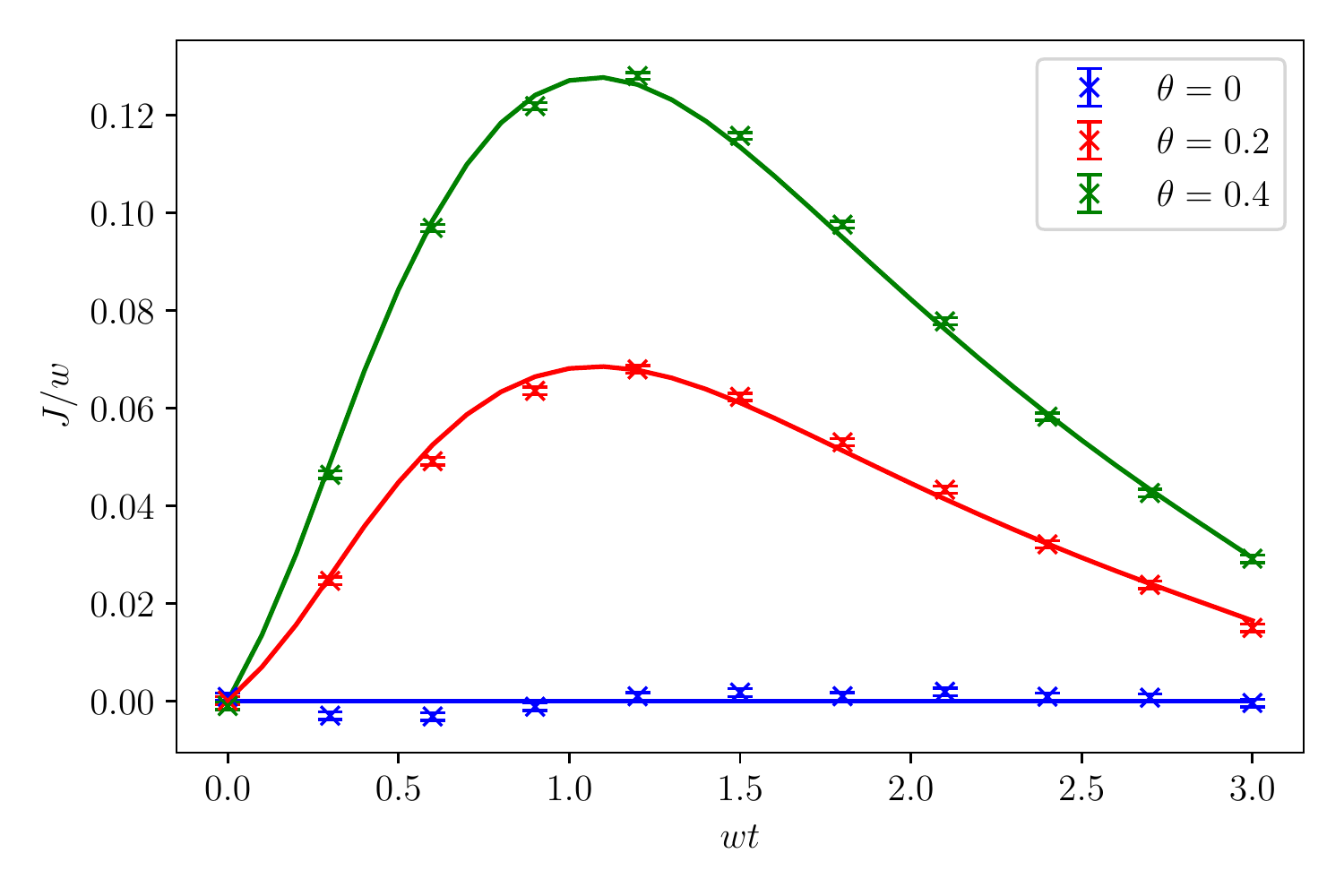}
\label{fig:current2b}
}\end{minipage}
\caption{The vector currents after the $\theta$ quench with lattice size $N=12$ and temporal lattice spacing $wt/S=0.3$. The fermion mass is set to be (a) $M=0.1w$ and (b) $M=0.5w$. The error bar for each data point represents the statistical error which are estimated by performing 100000 measurements. The solid lines are exact results.}
\label{fig:current2}
\end{figure}

Let us begin by discussing the features of our results at short time after the $\mu_5$ quench, where $\theta=-2\mu_5 t$:
\begin{enumerate}
\item
The expectation value of the vector current at short times \eqref{short} is given by
\begin{align}
\label{eq:short_mu5}
 &\langle\bar{J} \rangle
 = 4\mu_5mt^2 \langle\bar{\psi}\psi\rangle_\text{ave}+ \calO(t^3)
 \approx -c\mu_5m^2t^2,
\end{align}
where $\langle\bar{\psi}\psi\rangle_\text{ave}$ is the spatial average of the scalar condensate (proportional to $-m$), and $c$ is a constant.
The above expression for $\langle\bar{J} \rangle$ exhibits an approximately quadratic dependence on time after the quench, which agrees with the behavior seen in Figs \ref{fig:current1}.
\item The vector current is also approximately quadratic in the fermion mass $m$ and linear in the chiral chemical potential $\mu_5$, as expected from \eqref{eq:short_mu5}.
The explicit comparison between the data and the short-time behavior~\eqref{eq:short_mu5} is shown in Appendix~\ref{app:commutation}.
\end{enumerate}

Our quantum simulation results are in good agreement with the results based on exact diagonalization as shown in Figs.~\ref{fig:current1} and \ref{fig:current2}.
Note that the error bars account for the statistical error but not the Trotter error, that results in deviation from exact result as explained at the end of section \ref{subsec:evo}.
At larger times, the vector current exhibits a rich dynamics with non-linear dependence on time. Some features of this dynamics can still be understood qualitatively. 
For the $\mu_5$ quench, the current tends to a finite value at late times (see Fig.~\ref{fig:current1}), and then, at larger value of the mass, starts to exhibit saturation caused by the relaxation of chirality (see Fig.~\ref{fig:current1b}). For the $\theta$ quench, the current after the initial pulse at $t=0$ relaxes back to zero (see Fig.~\ref{fig:current2}), possibly with subsequent oscillations.  This relaxation is faster for a larger value of the fermion mass, in accord with (\ref{axial}). For larger fermion masses, we observe the oscillation; for smaller masses, the oscillations, if present, are characterized by a larger period.

The real-time dynamics following the $\theta$-quench was extensively studied in \cite{Zache:2018cqq} using exact diagonalization. 
The $\mu_5$-quench that we investigated here is directly relevant for the real-time dynamics of the chiral magnetic effect.

\section{Summary and outlook}

We have considered the behavior of a free model of relativistic fermions under the global {\it chiral quenches} that abruptly change the value of the $\theta$ angle. It has been observed here that such quenches induce vector currents stemming from the dynamics of chirality relaxation. The resulting real-time dynamics of the vector current appears unexpectedly rich, and is determined by the interplay between the processes of chirality pumping induced by the rotating $\theta$-angle and chirality absorption and relaxation.
\\
 
  This pilot study clarifies the effect of explicit breaking of chiral symmetry by fermion mass on the real-time dynamics of the Chiral Magnetic Effect. 
The computation that we have performed is amenable to a study using a classical computer; however, we feel that it is important to demonstrate the potential of quantum computation for describing the real-time dynamics of quantum field theories.
The study of the effect of fermion mass on real-time chiral dynamics is relevant for many practical applications - e.g. the  quarks in QCD, or chiral quasiparticles with a finite  gap in Dirac semimetals. In addition, the Schwinger model has been used to model the fragmentation of quarks in high-energy collisions \cite{casher1974vacuum,fujita1989quark,wong1991study,loshaj2012lpm,kharzeev2013jet,kharzeev2014anomalous,berges2018dynamics}, and the relaxation of chirality is key to the dynamics of this process. Our results show how the fermion mass affects the relaxation to a steady state, and the behavior of the vector current. \\

In the future, we plan to include the dynamical gauge field to investigate the interplay of anomalous and mass-induced chirality relaxation in the generation of the chiral magnetic current. This will require developing a method for truncating the $U(1)$ gauge field. For example, if $U(1)$ is approximated by $\bZ_N$, then in general per each site one needs $\log N$ qubits. However in Schwinger model, the absence of propagating gauge field modes allows to use $\log N$ qubits per entire lattice. This should allow to simulate the model with a moderate number of qubits.\\




\section*{Acknowledgement}

We would like to thank Bipasha Chakraborty, Masazumi Honda, Taku Izubuchi, and Akio Tomiya for important discussions on closely related works. We acknowledge the use of IBM Q quantum simulator. The work of D.K. was supported in part by the U.S. Department of Energy, Office of Nuclear Physics, under contracts DE-FG-88ER40388 and DE-AC02-98CH10886, and by the Office of Basic Energy Science under contract DE-SC-0017662.\\

\appendix
\section{Relation among Dirac fermion, staggered fermion, and spin operators}
\label{app:staggered}

Given a theory with two-component Dirac fermion $\psi(x)=(\rho(x), \eta(x))^\mt$ defined on a continuum spacetime, we place the theory on a spacial lattice by placing a one-component staggered fermion $\chi_n$ on each site. Letting lattice size be an even integer $N$ and lattice space be $a$, $\chi_{2m}/\sqrt{a}$ is identified with $\rho(x=m)$ and $\chi_{2m+1}/\sqrt{a}$ is identified with $\eta(x=m)$, where $m$ runs from $0$ to $N/2$.
With this correspondence in mind, the following operators are translated from Dirac fermion to staggered fermion as follows:
\begin{align}
\int\diff x\bar{\psi}\psi 
&\leftrightarrow
\sum_n (-1)^n\chi_n^\dag\chi_{n},
\\
\int\diff x\bar{\psi}\gamma_0\psi 
&\leftrightarrow
\sum_n \chi_n^\dag\chi_{n},
\\
\int\diff x\bar{\psi}\gamma_1\psi 
&\leftrightarrow
-\frac{1}{2}\sum_n \big[\chi_n^\dag\chi_{n+1}+\chi_{n+1}^\dag\chi_{n}\big],
\\
\int\diff x\bar{\psi}\gamma_5\psi 
&\leftrightarrow
-\frac{1}{2}\sum_n (-1)^n\big[\chi_n^\dag\chi_{n+1}-\chi_{n+1}^\dag\chi_{n}\big],
\\
\int\diff x\bar{\psi}\gamma_1\p_1\psi 
&\leftrightarrow
-\frac{1}{2a}\sum_n \big[\chi_n^\dag\chi_{n+1}-\chi_{n+1}^\dag\chi_{n}\big].
\end{align}
The left-hand sides correctly reproduce the right-hand sides in the continuum limit $a\to0$.
The operators written in terms of staggered fermions are further converted to spin operators by using the Jordan-Wigner transformation-\eqref{eq:JW}.

We summarize the relation of operators among these representations:
\begin{widetext}
\begin{align}
\begin{array}{c|c|c}
  \text{Dirac} & \text{staggered} & \text{spin}
 \\[1mm]\hline && \\
  \bar{\psi}\psi  & \displaystyle{\frac{(-1)^n}{a}\chi_n^\dag\chi_{n}} & \displaystyle{\frac{(-1)^n}{2a}Z_n}
 \\&&\\\hline && \\
  \bar{\psi}\gamma_0\psi  & \displaystyle{\frac{1}{a}\chi_n^\dag\chi_{n}} & \displaystyle{\frac{1}{2a}Z_n}
 \\&&\\\hline && \\
  \bar{\psi}\gamma_1\psi  & \displaystyle{\frac{1}{2a} \big[\chi_n^\dag\chi_{n+1}+\chi_{n+1}^\dag\chi_{n}\big]} &  \displaystyle{\frac{1}{4a}\big[X_nY_{n+1}-X_nY_{n+1}\big]}
 \\&&\\\hline && \\
  \bar{\psi}\gamma_5\psi  & \displaystyle{\frac{(-1)^{n}}{2a} \big[\chi_n^\dag\chi_{n+1}-\chi_{n+1}^\dag\chi_{n}\big]} &  \displaystyle{-\frac{\im(-1)^{n}}{4a}\big[X_nX_{n+1}+Y_nY_{n+1}\big]}
 \\&&\\\hline && \\
  \bar{\psi}\gamma_1\p_1\psi  & \displaystyle{-\frac{1}{2a^2} \big[\chi_n^\dag\chi_{n+1}-\chi_{n+1}^\dag\chi_{n}\big]} & \displaystyle{-\frac{\im}{4a^2}\big[X_nX_{n+1}+Y_nY_{n+1}\big]}
 \\&&
\end{array}
\end{align}
\end{widetext}
The expressions in each row are different representations of the same operators. More precisely, the operators in terms of staggered fermions and spins are equivalent up to constant terms and they agree with the operators of Dirac fermions in the continuum limit.
Note that for $n=N-1$ the operator $(-1)^{\frac{N}{2}+1}\prod_{i=1}^{N-2}Z_i$ has to be inserted in the spin representation in order to take account of the periodic boundary condition.

\section{Current at short time}
\label{app:commutation}

The vector current at shot time is given by,
\begin{widetext}
\begin{align}
 &\langle \bar{\calT}[\e^{\im \int_0^t\diff t' H}\big]J\,\calT[\e^{-\im\int_0^t\diff t' H}\big]\rangle_0
 =\im \int_0^t\diff t_1\langle [H(t_1),J]\rangle_0 -\int_0^t\diff t_1\int_0^{t_1}\diff t_2\langle [H(t_2),[H(t_1),J]]\rangle_0
 + \calO(t^3).
\end{align}
The commutators are calculated as,
\begin{align}
 [H(t_1),J] 
 &= \int\diff x\diff y\big[m\cos\theta(t_1)\bar{\psi}\psi(x) +\im m\sin\theta(t_1)\bar{\psi}\gamma_5\psi(x),\bar{\psi}\gamma_1\psi(y)\big]
 \nonumber\\
 &= \int\diff x\big(2m\cos\theta(t_1)\bar{\psi}\gamma_5\psi+2\im m\sin\theta(t_1)\bar{\psi}\psi\big),
 \\
 [H(t_2),[H(t_1),J]] 
 &= \int\diff x\diff y
 \big[
 \im\bar{\psi}\gamma_1\p_1\psi(x)-\frac{\dot{\theta}(t_2)}{2}\bar{\psi}\gamma_1\psi(x)+m\cos\theta(t_2)\bar{\psi}\psi(x)+\im m\sin\theta(t_2)\bar{\psi}\gamma_5\psi(x),
 \nonumber\\
 &\phantom{2m\cos\theta\bar{\psi}\gamma_5\psi(y)}
 2m\cos\theta(t_1)\bar{\psi}\gamma_5\psi(y)+2\im m\sin\theta(t_1)\bar{\psi}\psi(y)\big]
 \nonumber\\
 &= \int\diff x\big(-4\im m\cos\theta(t_1)\bar{\psi}\p_1\psi 
 + 4m\sin\theta(t_1)\bar{\psi}\gamma_5\p_1\psi
 \nonumber\\
 &\phantom{\diff x\diff x}
 +2m\dot{\theta}(t_2)\cos\theta(t_1)\bar{\psi}\psi
 +2\im m\dot{\theta}(t_2)\sin\theta(t_1)\bar{\psi}\gamma_5\psi
 +4m^2\bar{\psi}\gamma_1\psi\big).
\end{align}
Since the vacuum before the quench is parity symmetric,
\begin{align}
\langle\bar{\psi}\gamma_5\psi\rangle_0 = 0,
\qquad
\langle\bar{\psi}\gamma_1\psi\rangle_0 = 0,
\qquad
\langle\bar{\psi}\p_1\psi\rangle_0 = 0.
\end{align}
\begin{figure}[t]
\centering
\includegraphics[scale=0.5]{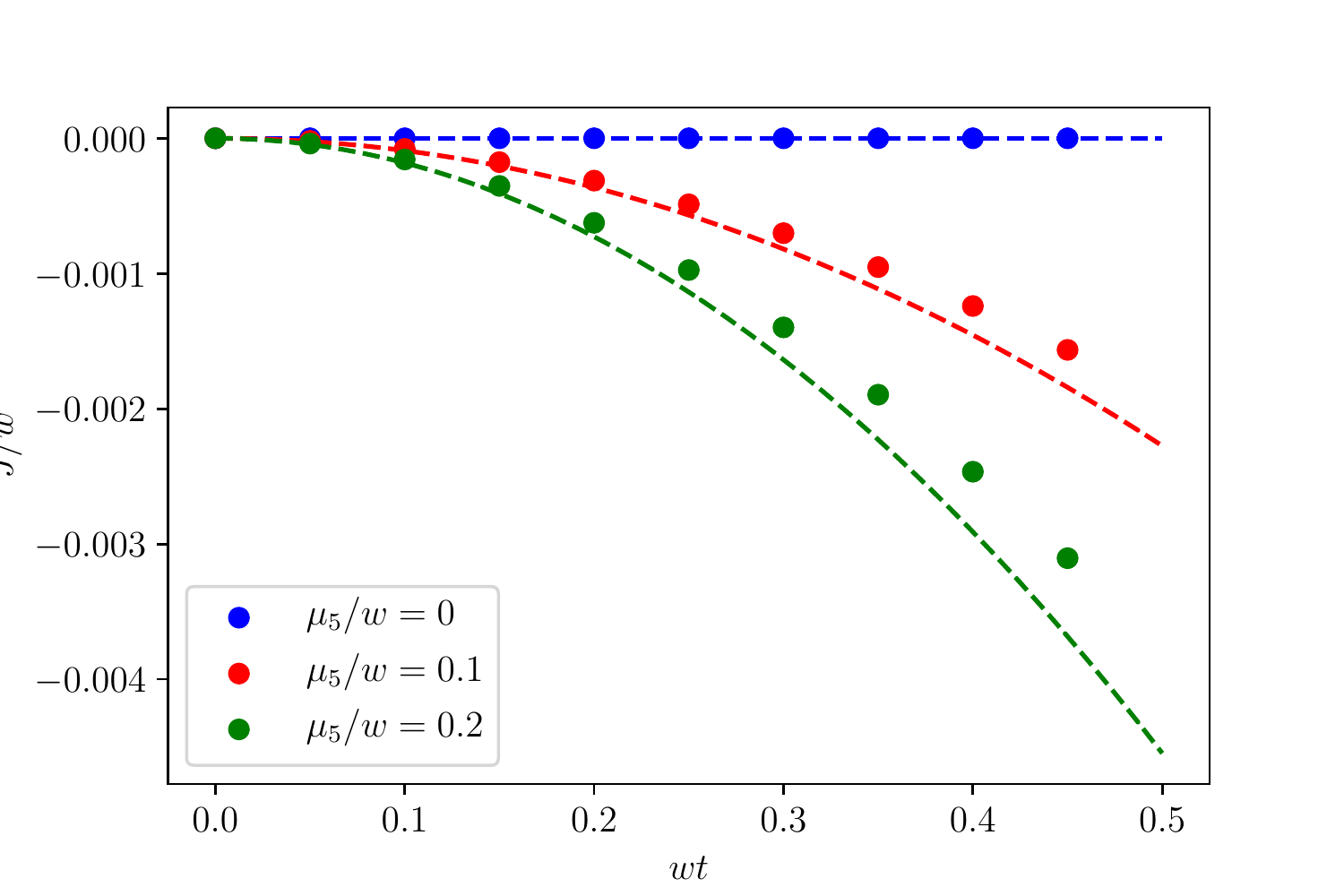}
\caption{Short-time behavior of the vector current following the $\theta$ quench with lattice size $N=12$ and fermion mass $M=0.1w$. The data points are obtained by exact diagonalization and dashed lines are short-time behavior of the current~\eqref{eq:current_short}. }
\label{fig:shortT}
\end{figure}
Hence, the current is computed as
\begin{align}
 \langle \bar{\calT}[\e^{\im \int_0^t \diff t' H}]J\, \calT[\e^{-\im \int_0^t\diff t'H}]\rangle_0
 &=\Big(-2m\int_0^t\diff t_1\sin\theta(t_1)
 -2m\int_0^t\diff t_1\int_0^{t_1}\diff t_2\dot{\theta}(t_2)\cos\theta(t_1)\Big)
 \int\diff x\langle\bar{\psi}\psi\rangle_0
 \nonumber\\
 &-4m\int_0^t\diff t_1\int_0^{t_1}\diff t_2\sin\theta(t_1)\int\diff x\langle\bar{\psi}\gamma_5\p_1\psi\rangle_0
 + \calO(t^3).
\end{align}
\end{widetext}
In case of the $\mu_5$-quench, where $\theta=-2\mu_5 t$, it is further simplified to
\begin{align}
\label{eq:current_short}
 \langle \bar{\calT}[\e^{\im \int_0^t\diff t'H}]J\, \calT[\e^{-\im\int_0^t\diff t' H}]\rangle_0
 &=4m\mu_5t^2\int\diff x\langle\bar{\psi}\psi\rangle_0
 + \calO(t^3).
\end{align}
The short-time behavior is shown in Fig.~\ref{fig:shortT}.

\section{Quantum circuits for time evolution}
\label{app:circuit}

We show the circuit implementations used for time-evolution operator~\eqref{eq:suzuki_trotter}.

\begin{align}
 \e^{-\im\alpha Z_iZ_j}=\qquad
\begin{array}{c}
 \Qcircuit @C=2mm @R=2mm{
 \lstick{j} & \targ & \gate{R_Z(2\alpha)} & \targ &\qw
 \\
 \lstick{i} & \ctrl{-1} & \qw & \ctrl{-1} &\qw
 }
\end{array}
\end{align}
where $R_{Z}(\theta):=\e^{-\im\frac{\theta}{2}Z}$.
\begin{align}
 \e^{-\im\alpha (X_i X_j+Y_i Y_j)}=\qquad
\begin{array}{c}
 \Qcircuit @C=2mm @R=2mm{
 \lstick{j} &\targ    &\qw       &\targ     &\gate{R_Z(-2\alpha)}&\targ    &\qw       &\targ    &\qw
 \\
 \lstick{i} &\ctrl{-1}&\gate{H}&\ctrl{-1}&\gate{R_Z(2\alpha)}&\ctrl{-1}&\gate{H}&\ctrl{-1}&\qw
 }
\end{array}
\end{align}
Then, $\e^{-\im\alpha (XY-YX)}$ is immediately implemented by noting
\begin{align}
 \e^{-\im\alpha (X_iY_j-Y_iX_j)} = S_j\e^{-\im\alpha (X_iX_j+Y_iY_j)}S^\dag_j.
\end{align}
The evolution operator involving a nonlocal term is implemented by the following circuit,
\begin{align}
 \e^{-\im\alpha (X_0 X_{N-1}+Y_0Y_{N-1})\prod_{i=1}^{N-2}Z_i}=\qquad\qquad
\begin{array}{c}
 \Qcircuit @C=2mm @R=2mm{
 \lstick{N-1} &\targ    &\qw&\qw&\qw&\cdots&&\qw       &\targ     &\gate{R_Z(-2\alpha)}
 &\targ&\qw    &\qw&\cdots&&\qw&\qw&\targ    &\qw
 \\
 \lstick{N-2} &\qw   &\qw&\qw&\qw&\cdots&&\ctrl{6}       &\qw     &\qw
 &\qw &\ctrl{6}&\qw&\cdots&&\qw&\qw&\qw    &\qw
 \\
 \\
 \lstick{\vdots} &&&&& \udots &&&&&&&& \ddots
 \\
 \\
 \\
 \lstick{1} &\qw   &\qw&\ctrl{1}&\qw&\cdots&&\qw&\qw       &\qw     
 &\qw&\qw&\qw&\cdots&&\ctrl{1}&\qw       &\qw    &\qw
 \\
 \lstick{0} &\ctrl{-7}&\gate{H}&\targ&\qw&\cdots&&\targ&\ctrl{-7}&\gate{R_Z(2\alpha)}
 &\ctrl{-7}&\targ &\qw&\cdots&&\targ&\gate{H}&\ctrl{-7}&\qw
 }
\end{array}
\end{align}
Then, $\e^{-\im\alpha (XY-YX)}$ is immediately implemented by noting
\begin{align}
 \e^{-\im\alpha (X_0Y_{N-1}-Y_0X_{N-1})} 
 = S_{N-1}\e^{-\im\alpha (X_0 X_{N-1}+Y_0Y_{N-1})\prod_{i=1}^{N-2}Z_i}S^\dag_{N-1}.
\end{align}

\bibliographystyle{utphys}

\bibliography{QIS}
\end{document}